\documentclass{aa}
\usepackage{graphicx}
\usepackage{natbib}

\begin{document}

\newcommand{\CC}{$^{12}$C$^{12}$C}
\newcommand{\CCC}{$^{12}$C$^{13}$C}
\newcommand{\CCCC}{$^{13}$C$^{13}$C}
\newcommand{\CN}{$^{12}$C$^{14}$N}
\newcommand{\CCN}{$^{13}$C$^{14}$N}
\newcommand{\tm}{$<\tau_{1{\mu}m}>$]}
\newcommand{\ta}{$\tau_{1{\mu}m}$}
\newcommand{\CO}{$^{12}$C$^{16}$O~}
\newcommand{\CCO}{$^{13}$C$^{16}$O~}
\newcommand{\CDC}{$^{12}$C/$^{13}$C}

\newcommand{\Vm}{$V_{\rm macro}$}
\newcommand{\Vt}{$V_{\rm t}$}
\newcommand{\Vexp}{$V_{\rm exp}$}
\newcommand{\Vmacro}{$V_{\rm macro}$}
\newcommand{\Vr}{$V_{\rm rotat}$}

\newcommand{\HHO}{H$_2$O}
\newcommand{\DG}{$\Delta_t$}

\newcommand{\Tef}{\mbox{$T_{\rm eff}$}}
\newcommand{\Msun}{\mbox{\,M$_\odot$}}
\newcommand{\Lsun}{\mbox{\,L$_\odot$}}
\newcommand{\vunit}{\mbox{\,km\,s$^{-1}$}}
\newcommand{\mic}{\mbox{$\,\mu$m}}
\newcommand{\pion}[2]{{#1}\,{\sc {#2}}}
\newcommand{\fion}[2]{[{#1}\,{\sc {#2}}]}
\newcommand{\ltsimeq}{\raisebox{-0.6ex}{$\,\stackrel
        {\raisebox{-.2ex}{$\textstyle <$}}{\sim}\,$}}
\newcommand{\gtsimeq}{\raisebox{-0.6ex}{$\,\stackrel
        {\raisebox{-.2ex}{$\textstyle >$}}{\sim}\,$}}
\newcommand{\astrut}{\mbox{\rule[-4mm]{0mm}{8mm}}}

\title{CO bands in V4334~Sgr (Sakurai's Object): the $^{12}$C/$^{13}$C ratio}

\author{Ya. V. Pavlenko\inst{1}, T. R. Geballe\inst{2}, A. 
Evans\inst{3}, B. Smalley\inst{3}, S. P. S. Eyres\inst{4},
V. H. Tyne\inst{3}, L.A. Yakovina\inst{1}
}

\offprints{Ya. V. Pavlenko}
\mail{yp@mao.kiev.ua}

\institute{Main Astronomical Observatory, Academy of Sciences of the Ukraine, Golosiiv
     Woods, Kyiv-127, 03680 Ukraine
\and Gemini Observatory, 670 N. A'ohoku Place, Hilo, Hi 96720, USA
\and Astrophysics Group, School of Chemistry \& Physics, Keele University, Keele
     Staffordshire, ST5 5BG, UK
\and Centre for Astrophysics, University of Central Lancashire, Preston, Lancashire
     PR1 2HE, UK }

\date{January 17, 2004}

\authorrunning{Pavlenko et al.}
\titlerunning{$^{12}$C/$^{13}$C in V4334 Sgr}

\abstract{We present the results of our analysis of a high resolution
($R\simeq30000$) infrared spectrum of V4334~Sgr (Sakurai's Object)
around the first overtone CO bands, obtained in 1998 July. The $^{12}$CO
and $^{13}$CO bands are well-resolved, and we compute synthetic
hydrogen-deficient model atmosphere spectra to determine the \CDC\ ratio.
We find \CDC\ $\simeq4\pm1$, consistent with the interpretation of V4334~Sgr
as an object that has undergone a very late thermal pulse. 
 }

\maketitle

\section{Introduction} V4334~Sgr (Sakurai's Object) is a low
mass ($\sim0.6$-$0.8\Msun$; \citealt{herwig}) star undergoing a
very late thermal pulse (VLTP). After traversing the `knee' on
the post-asymptotic giant branch (AGB) evolutionary track, the
star experiences a final He flash as it evolves down the white
dwarf cooling track: it becomes a `born-again' AGB star (see
\citealt{herwig, lawlor}).

A key factor in confirming this scenario is the determination of
elemental abundances and isotopic ratios in the surface
material. \cite{asplund99} have used optical spectroscopy to
determine the abundance of a range of species; they found
V4334~Sgr to be hydrogen-deficient, with other elemental
abundances close to those of the R~CrB stars; 
R~CrB stars. They also used the C$_2$ (1-0) and 
(0-1) Swan bands at 4740 and 5635\AA\ respectively to determine that the 
\CDC\ ratio was somewhere in the range 1.5--5, although they did not give 
an actual value.

\cite{pav-trg} modelled the spectral energy distribution of V4334~Sgr
in the near-IR ($\lambda\lambda$ 1 - 2.5 \mic). They broadly
confirmed the results of \cite{asplund99}, but found some 
evidences of higher C abundance, by a factor $\sim2$.
\citeauthor{pav-trg} also noted the presence of hot dust longward
of $\sim2\mic$.

The \CDC\ ratio is of fundamental importance in understanding
the VLTP process, as it is a reflection of the relative
importance of $^{12}$C, $^{13}$C processing and dredge-up. In
this Letter we use echelle observations of the first overtone
vibration-rotation features of CO, obtained in 1998 July while
the star was still visible, to determine the \CDC\ ratio in
V4334~Sgr.

\section{Observation}

The data were obtained with the facility spectrometer CGS4
\citep{mountain} at the 3.8m United Kingdom Infrared
Telescope on UT 1998 July 7. CGS4 was configured with a
$256\times256$ InSb array, its 31~l\,mm$^{-1}$ echelle and
a 0.45\arcsec\ slit, which provided a resolving power of
$\sim30,000$. The spectrum was sampled every 0.4 resolution
element. Five echelle settings were used to continuously cover
the wavelength range 2.32--2.38\mic, an interval containing
numerous $^{12}$CO and $^{13}$CO lines, including three bandheads
of $^{12}$CO ($\delta\upsilon= 3\rightarrow1, 4\rightarrow2,
5\rightarrow3$) and two of $^{13}$CO ($\delta\upsilon=2\rightarrow0,
3\rightarrow1$). Each of the five spectral segments
was observed for a total of 8 minutes, with individual data points
being observed for 4 minutes. A spectrum of the A2V star HR6378
was obtained just prior to each spectrum of V4334~Sgr and at
similar airmass. The raw spectra of V4334~Sgr and HR6378 were
extracted, reduced, and ratioed using standard techniques.
Wavelength calibration was achieved using some of the multitude
of telluric absorption lines in the spectra of the calibration
star, and is accurate to $\pm3$\vunit; the wavelength scale is
in vacuum. The ratioed spectral segments slightly overlap one
another and were scaled and adjoined to provide the final spectrum
used in the analysis.

\section{Procedure}

\subsection{Spectral synthesis}
We have used the technique of synthetic spectra to carry out
our analysis of the IR spectrum of V4334~Sgr. We computed the
spectra in the classical framework, assuming LTE, plane-parallel
media, and no sinks and sources of energy in the atmosphere. The
transfer of energy is provided by the radiation field and by
convection. Strictly speaking, none of these assumptions is
completely valid in the atmosphere of V4334~Sgr because, in
all probability, we see only the pseudo\-photosphere, i.e.
the outermost part of the expanding envelope. However, we
suggest that none of these assumptions is of crucial importance
for the spectrum formation processes.

A grid of LTE synthetic spectra was computed for \CDC\ ratios 
of 1, 2, 3, 4, 5, 6, 7, 8, 9, 10,  model atmosphere parameters 
$\Tef,\log g$  = 5250,0.0 \citep{pav-hwd}, in the wavelength
range 2.31--2.39\mic\ and with wavelength step 0.00002\mic,
using the WITA612 code \citep{pavlenko}. We used abundances
as given by \cite{asplund99}, scaled such that $\sum N_i=1$,
and with N, O abundances of --2.52, --2.02 respectively, and
two abundances of carbon, namely $\log N(C) = -1.05$ and
$\log N(C) = -1.62$; the first case corresponds to a
carbon-rich model atmosphere (see \citealt{asplund99} for a
discussion of the carbon problem). As the determination of
the \CDC\ ratio in stellar atmospheres depends on \Vt\
(see \citealt{pavlenko03b}), we determined \CDC\ for two
values of the microturbulent velocity, \Vt = 3 and 6\vunit.

A set of continuum opacity sources from ATLAS9  
\citep{kurucz} was used. Additionally, 
all the major sources of opacity in our spectral region of interest
were taken into account:
\begin{enumerate}
\item [(i)] bound-free absorption by \pion{C}{i}, \pion{N}{i},
      \pion{O}{i} atoms computed by \cite{pavlenko03} for the
      TOPBASE cross-sections \citep{seaton};
\item [(ii)] \CO, \CCO\ lines from \cite{goorvitch};
\item [(iii)] \CC, \CCC, \CCCC, \CN\ and \CCN\ lines from
      \cite{kurucz}; 
\item [(iv)] atomic lines from VALD \citep{kupka}.
\end{enumerate}

\begin{figure}
\begin{center}
\includegraphics [width=88mm]{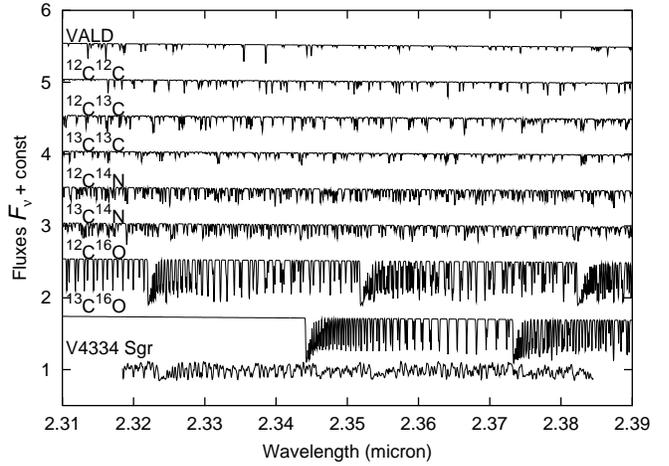}
\end{center}
\caption[]{\label{contrib} Theoretical spectra of different
species in the spectra region convolved by gaussian of
half-width 5$\times10^{-6}$\mic. Synthetic spectra were computed for 
\CDC =1}
\end{figure}

\begin{figure*}
\setlength{\unitlength}{1cm}
\begin{picture}(8.0,5.5)
\put(-0.3,-2.0){\includegraphics{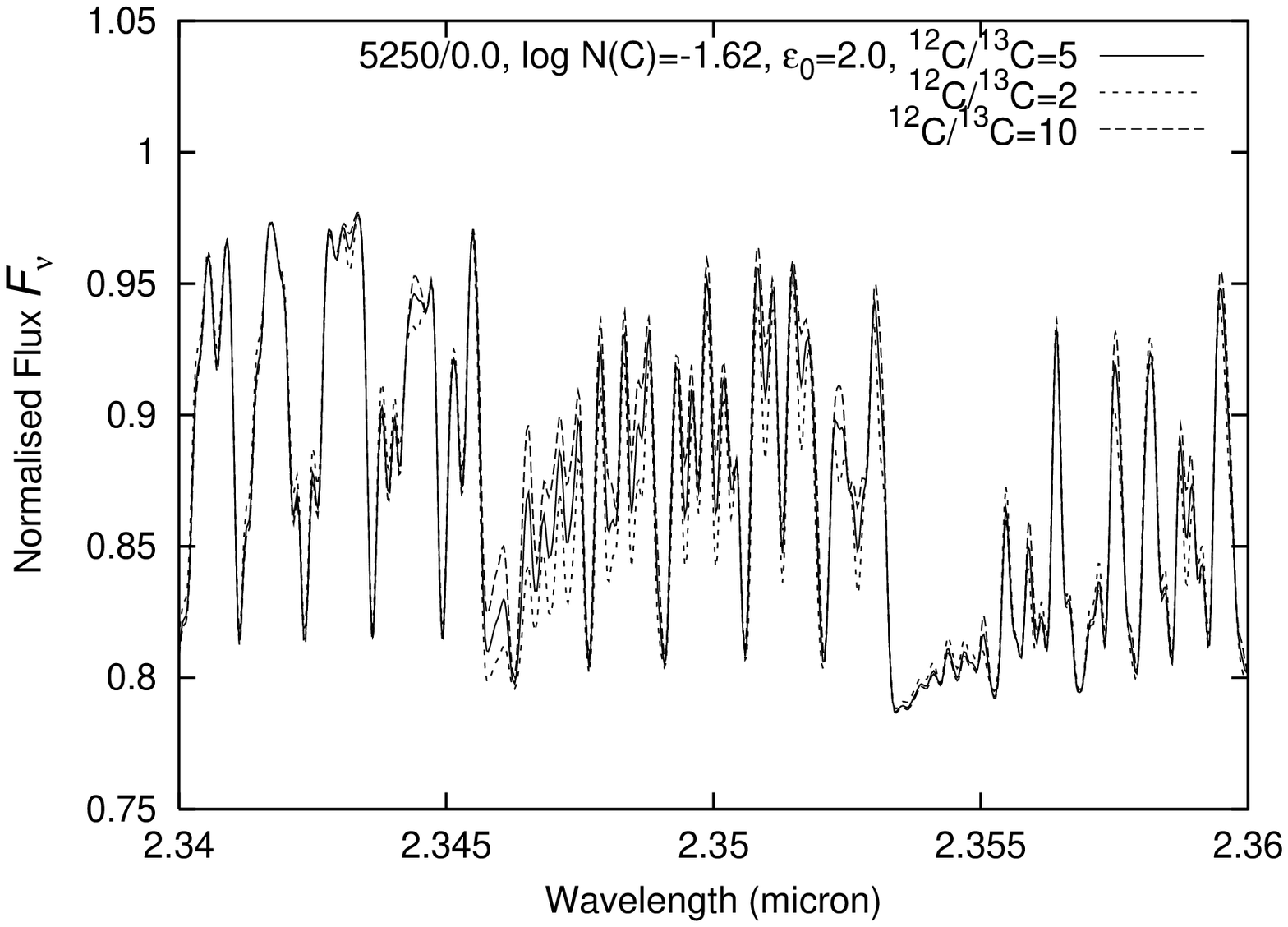}}
\put(-0.3,-2.0){\includegraphics{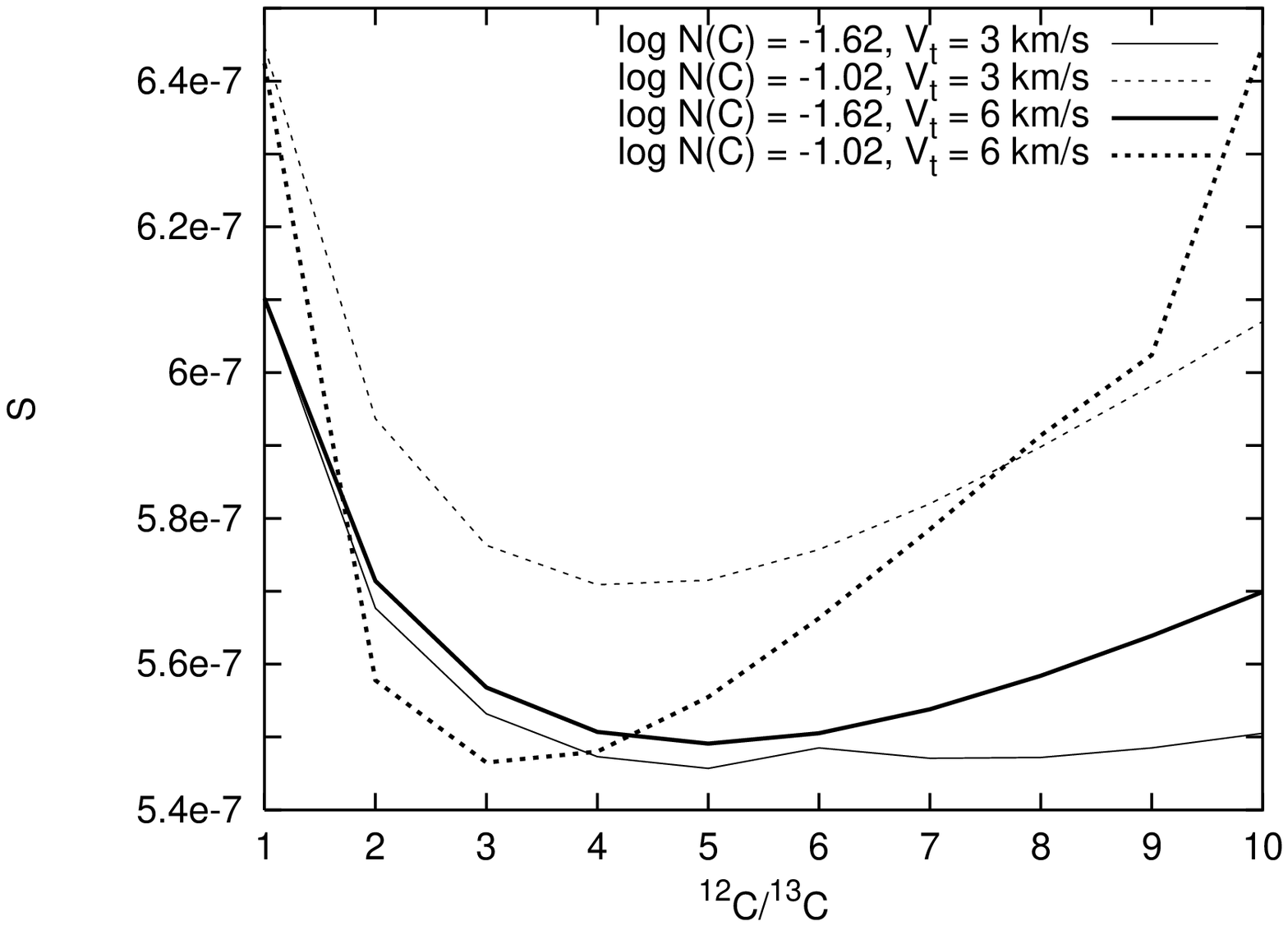}}
\end{picture}
\caption{ {\bf a)} Dependence of computed spectra on \CDC\ 
for values of \Tef, $\log g$ and $\epsilon_0$ indicated.
{\bf b)} Dependence of the parameter $S$ on \CDC\
for values of $\log N(\mbox{C})$ and \Vt\ indicated.
\label{__s}} 
\end{figure*}

We adopt the Voigt profile $H(a,v)$ for the shape of each
line; damping constants $a = (\gamma_2 +
\gamma_4 + \gamma_6)/(4\times\pi\times\Delta\nu_D)$ were
computed using data from various databases \citep{kurucz,kupka},
or computed following an approximation of \cite{unsold}. The
contributions of the various mole\-cular and atomic species to
the total opacity in the 2.31--2.39\mic\ spectral region are
shown in Fig.~\ref{contrib}, from which we see that absorption
by CO dominates. Finally, bearing in mind the high luminosity
of V4334~Sgr ($\sim10^4$\Lsun; \citealt{herwig}), we adopt
$\log g = 0$.

\subsection{The effect of circumstellar dust}

\cite{pav-trg} have shown that the 2\mic\ spectral region of
V4334~Sgr is affected by emission from the hot dust known to
envelop V4334~Sgr at this time, and which most likely formed
at some 20 stellar radii from the stellar atmosphere
\citep{kipper, tyne02}. The dust shell first appeared in 1997
March and, by 1998 August, its presence was well established
photometrically \citep{kerber,kipper} and spectroscopically
\citep{tyne00,tyne02}. 

The dust emission veils the CO absorption lines, and to
estimate its effect we suppose that the dust emission adds
some continuum flux in our spectral region, effectively
reducing the residual fluxes
$r_{\nu} = F_{\nu}^{\rm l+c} / F_{\nu}^{\rm c}$
(and hence equivalent widths) of the absorption lines;
here $F_{\nu}^{\rm l+c}$ and $F_{\nu}^{\rm c}$ are fluxes in
the lines and continuum respectively. We describe this
by the factor
$$r_{\nu} = (r_{\nu}^0 + \epsilon_0)/(1 + \epsilon_0), $$
where $r_{\nu}^0$ and $\epsilon_0$ are respectively the residual
flux computed for the `dust-free' case, and the relative contribution
of the dust emission; the parameter $\epsilon_0$ is determined from
comparison of computed and observed spectra.

\subsection{Fitting procedure}

To determine the best fit parameters, we compare the observed
residual fluxes $r_{\nu}$ with computed values
following the scheme of \cite{jones} and \cite{pav-jones}. We
let 
$$r_{\nu}^x = \int r^y_{\nu} \times G (x-y)*dy, $$
where $r^y_{\nu}$ and $G(x-y)$ are respectively the residual
fluxes computed by WITA612 and the broadening profile; we adopt
a gaussian for the latter. We then find the minima of the 3D
function 
 $$S(f_{\rm s}, f_{\rm h}, f_{\rm g}) = 
   \sum \left ( 1 - f_{\rm h} \times r^{\rm synt}/r^{\rm obs}  \right 
   )^2  , $$
where $f_{\rm s}$, $f_{\rm g}$, $f_{\rm g}$ are respectively the
wavelength shift, normalisation factor, and half-width of
$G(x-y)$. The parameters $f_{\rm s}, f_{\rm h}$ and $f_{\rm g}$
are determined by the minimization procedure for every computed
spectrum. Then, from the grid of the better solutions for a given
\CDC\ and $\epsilon_0$, we choose the best-fitting solution. 

\section{Results}
\label{results}


\begin{table}
\caption {\label{table1} Atmospheric and dust parameters for
V4334~Sgr in 1998 July.}
\begin {tabular} {lcccc}
\hline
$\log N(\mbox{C})$                      & \multicolumn{2}{c}{--1.05} &  \multicolumn{2}{c}{--1.62} \\ 
\Vt\ ($\!$\vunit)                           & 3 & 6 & 3 & 6 \\ \hline
                                       &&         &&      \\
Shift ($10^{-4}$\mic) between &&         &&      \\
\multicolumn{1}{r}{observed and computed spectra}
                                   &   8.6  & 8.6   &  8.6  & 8.6  \\
Normalisaton factor $f_{\rm h}$    &   0.96 & 0.96  &  0.96 & 0.96 \\
FWHM  ($10^{-4}$\mic)              &  1.56  & 1.4   &  1.8  & 1.4  \\
$\epsilon_0$                       &  1.4   & 2.0   & 1.8   & 2.4  \\
\astrut \CDC                       &   4    & 3     & (5,8) & 5    \\
                                       \hline\hline
\end{tabular}
\end{table}

On July 7 the Earth's orbital motion redshifted the spectrum of 
V4334~Sgr by 8\vunit. Thus the shift of +112\vunit\ that we needed
to apply means that the heliocentric velocity of the photosperic
CO was $+104\pm3.2$\vunit, where the uncertainty includes
$\pm3$\vunit\ for the wavelength scale, and $\pm1$\vunit\ in our
shift. While this is close to the known heliocentric velocity
of V4334~Sgr, $+115\vunit$ \citep{duerbeck}, it may also imply that
the photospheric layers in which the CO was located were
expanding at $\sim10$\vunit.

Our results show that \CDC\ depends on several parameters. 
Fig.~\ref{__s}a shows the dependence of the synthetic spectra on 
the \CDC\ ratio for $\epsilon_0=2.0$, while Fig.~\ref{__s}b shows 
the dependence of the best-fit parameter $S$ on the \CDC\ ratio 
for a variety of $\log N(\mbox{C})$ and \Vt\ values. Both the
\CO\ and \CCO\ bands are strong in the spectra due to the high
carbon abundance; the computed central intensities of the
strongest lines are $\sim0.4$ of the normalized continuum.

\begin{figure*}
\includegraphics [width=176mm]{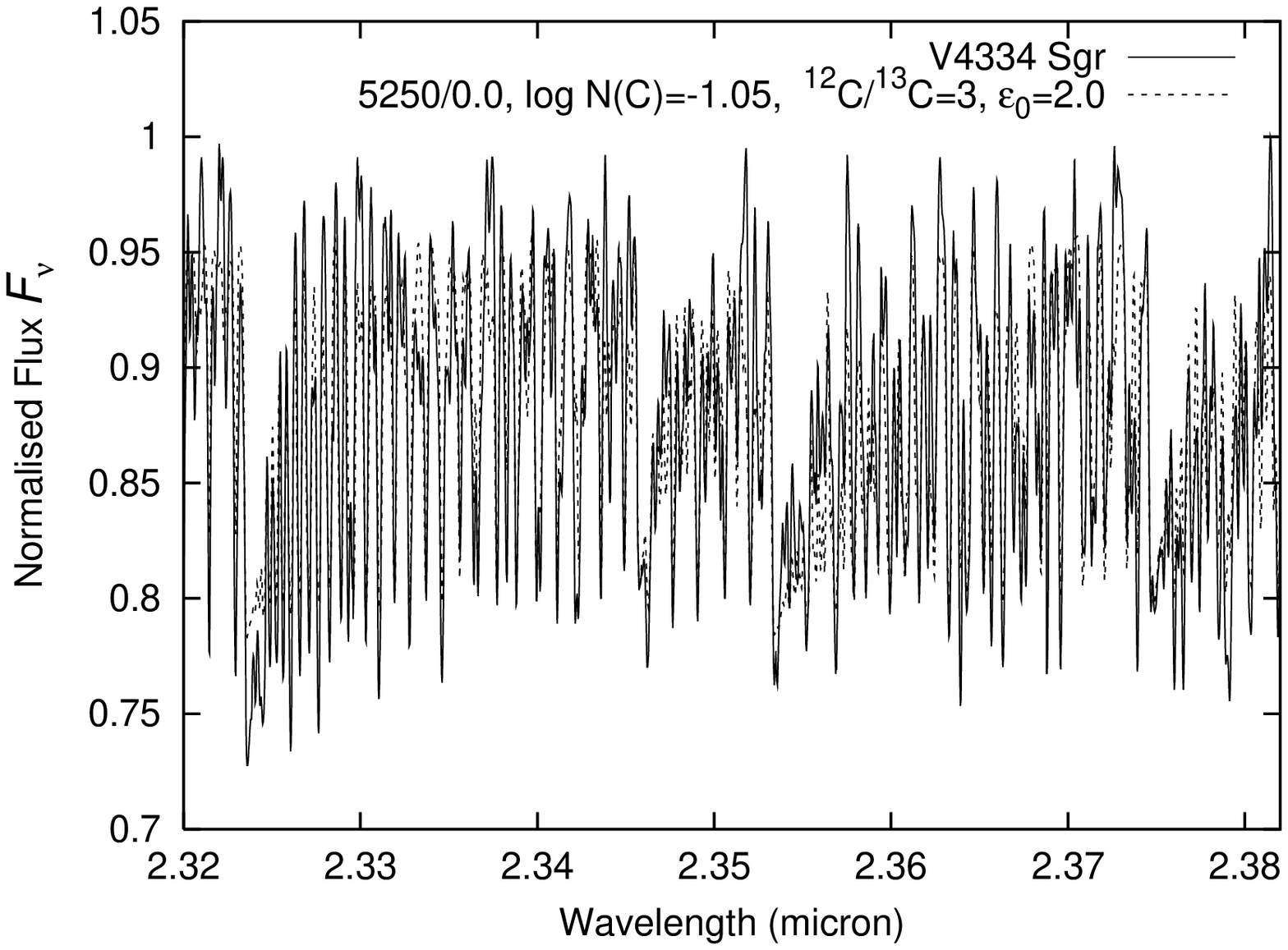}
\includegraphics [width=176mm]{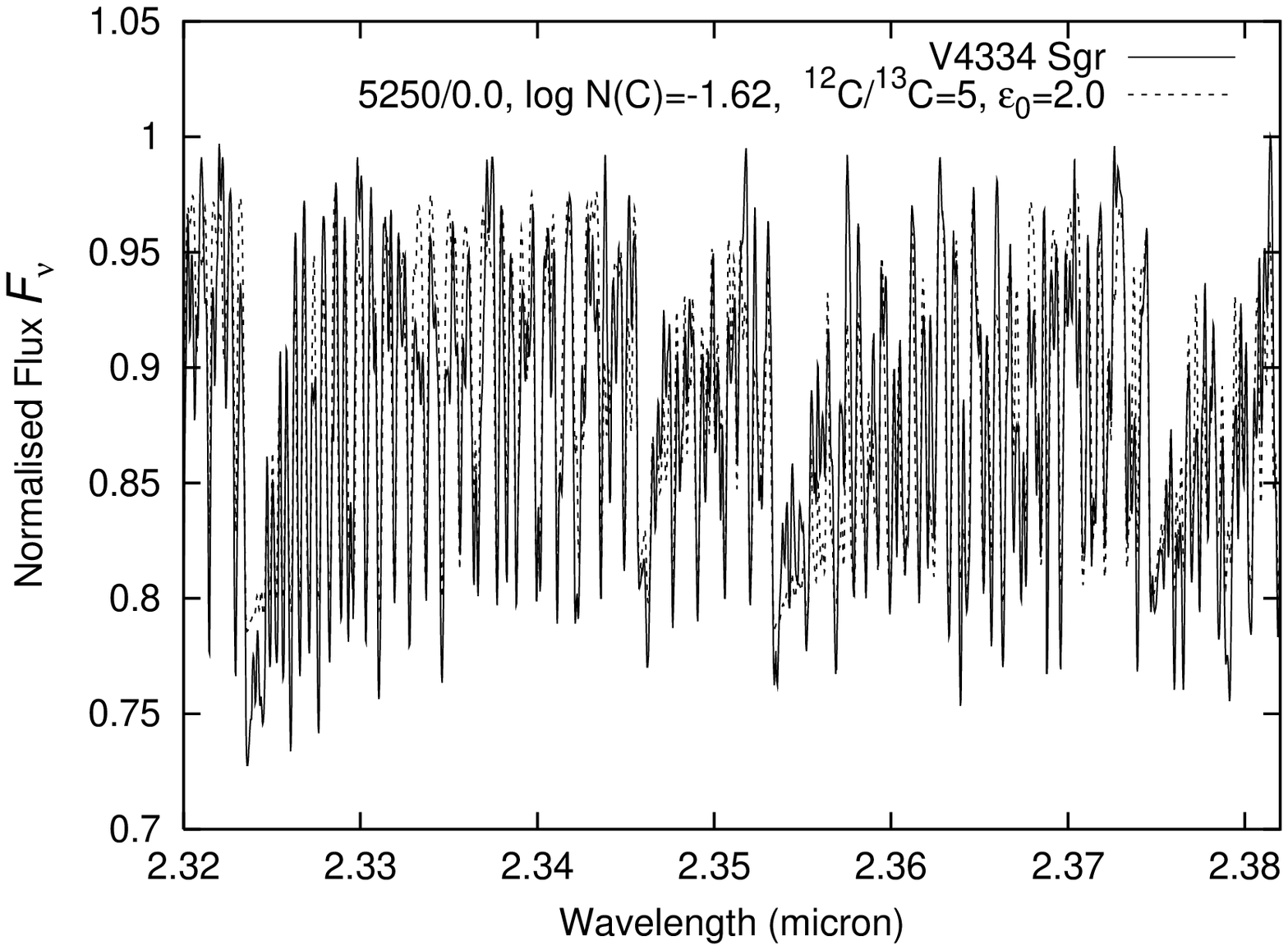}
\caption{{\bf a)} Best fit to 1998 July spectrum found by the minimisation
procedure outlined in the text, for $\log N(\mbox{C}) = -1.05$. {\bf b)} As
{\bf (a)}, but for $\log N(\mbox{C}) = -1.62$\label{06.fits}}
\end{figure*}

\begin{figure*}
\setlength{\unitlength}{1cm}
\begin{picture}(8.0,5.)
\put(-0.3,-2.0){\includegraphics{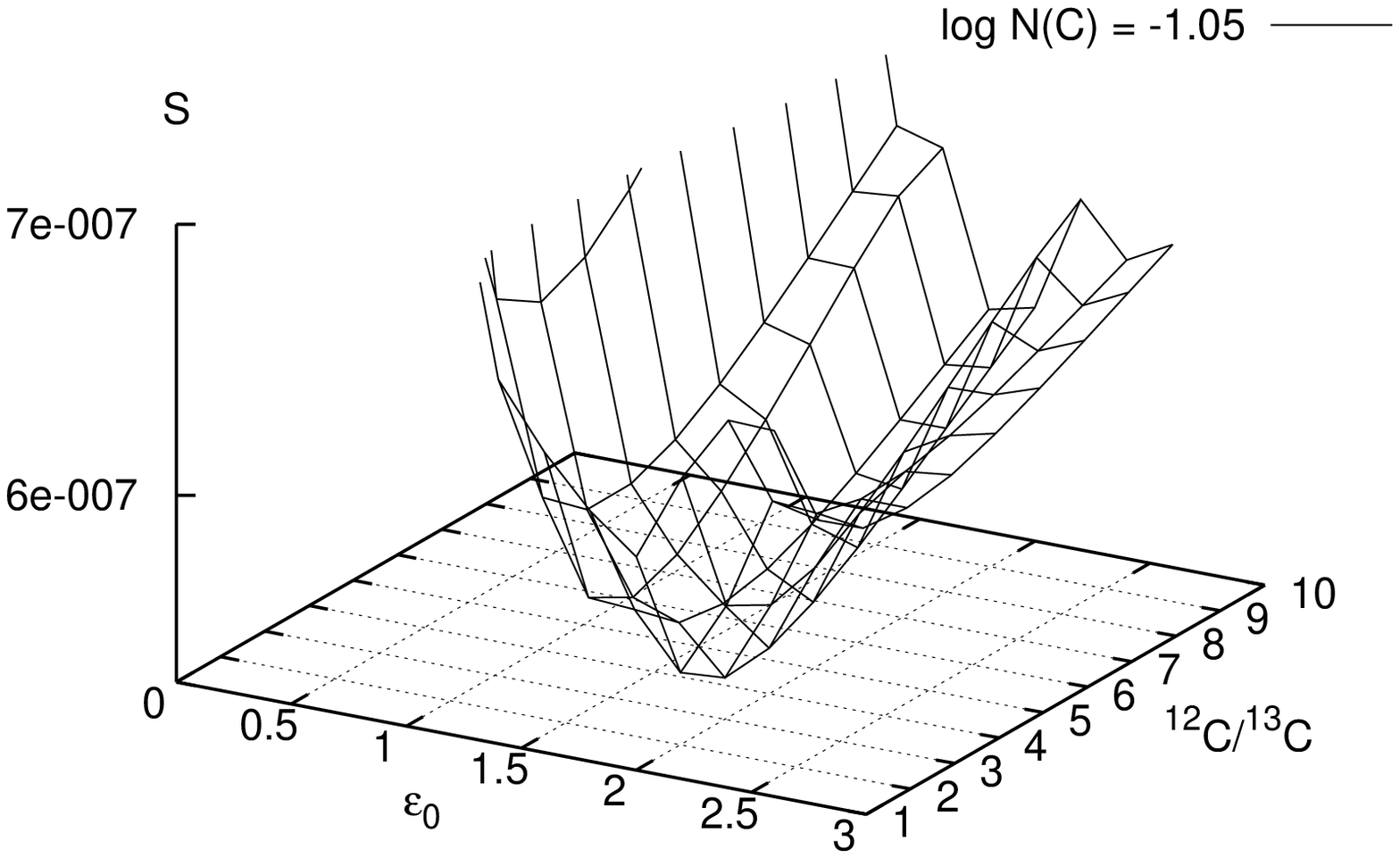}}
\put(-0.3,-2.0){\includegraphics{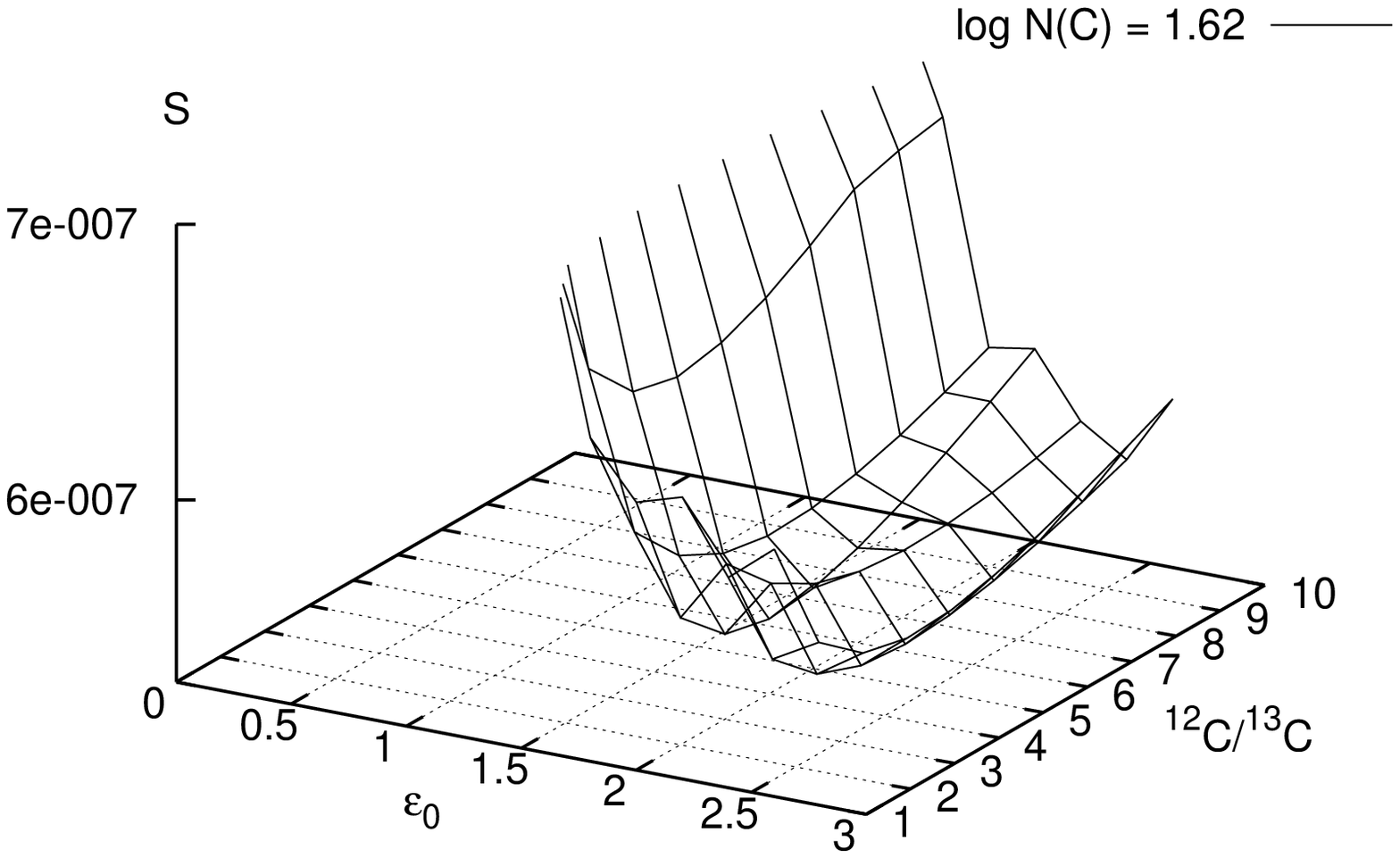}}
\end{picture}
\caption{{\bf a)} Dependence of the fit parameter $S$ on \CDC\ and $\epsilon_0$ for
a grid of the best fits found by minimisation procedure for $\log N(\mbox{C})
= -1.05$. {\bf b)} As (a), but for $\log  N(\mbox{C}) = -1.62$.}
\label{06.3d}
\end{figure*}

Using our fitting procedure we found a clear dependence of
the spectra on \CDC, \Vt\ and $\epsilon_0$. The best-fit synthetic
spectra for our two assumed C abundances are shown in
Figs.~\ref{06.fits}a,b; the corresponding parameters are given in
Table~\ref{table1}. Note that there are two possible solutions
for [$\log N(\mbox{C})$, \Vt] = [--1.62, 3] (see Fig.~\ref{__s}b).
Fig.~\ref{06.3d}a,b shows the 2D-dependence of the fit parameter
$S$ on \CDC\ and $\epsilon_0$. We effectively found two families
of solutions, differing by their \CDC\ ratios and $\epsilon_0$
values. An alternative interpretation is that there is an uncertainty
in our results dependent on the uncertainties in the input parameters.

In general, a mictoturbulent velocity \Vt = 6\vunit\ is most
likely for a supergiant atmosphere and our results for \Vt = 3\vunit,
presented in Table~\ref{table1}, are included primarily to illustrate
the \Vt-dependence of our results. We see from Figs~\ref{__s}b and
\ref{06.3d}, and Table~\ref{table1}, that a \CDC\ ratio of $4\pm1$ is
consistent with the data.

\section{Discussion}
\label{Discuss}


One of the major challenges posed by V4334~Sgr has been the extreme
rapidity with which it has evolved following the VLTP. \cite{herwig} has
addressed this problem by supposing that the element mixing efficiency
in the He flash convection zone during the VLTP is much smaller than
predicted by mixing length theory. Herwig defines the parameter
$f=D_{\rm MLT}/D_{\rm CM}$, where $D_{\rm MLT}$ is the diffusion
coefficient for convective element mixing and $D_{\rm CM}$ that for
composition mixing. \cite{herwig} finds that the rapid evolution of
V4334~Sgr is accounted for by $f\sim10-100$, and that, for this value of
$f$, the surface \CDC\ ratio for the $f = 30$ VLTP model is $\sim5$.
This is clearly very much in line with the results presented here.

\cite{pav-trg} found that dust made a major contribution to the
emission longward of 2\mic\ in 1998, and indeed was present as
early as 1997. The value of the parameter $\epsilon_0$ gives an indication
of the level of the dust emission at 2.3\mic. From Table~\ref{table1}
we see that $\epsilon_0\simeq2\pm0.2$, indicating that the dust flux at
2.3\mic\ is $\simeq9.4\times10^{-12}$~W~m$^{-2}$\mic$^{-1}$. This is
much higher than in later spectra, when the star had become completely
obscured and the dust emission dominated the overall spectrum
\citep{tyne02}. At later times ($\gtsimeq1999$) the dust envelope
expanded and cooled and the peak of its spectrum shifted to longer
wavelengths.

\section{Conclusion}

We have carried out a spectral synthesis to determine the \CDC\ ratio
in V4334~Sgr. Given a range of C abundance, we determine that \CDC\
$\simeq4\pm1$, consistent with the VLTP interpretation of V4334~Sgr. The
relative dust contribution at 2.3\mic\ is $\epsilon_0\simeq2\pm0.2$.
Our value of the \CDC\ ratio is quite well constrained, and will
doubtless be useful in fine-tuning the VLTP scenario.

\section*{Acknowledgments}
The United Kingdom Infrared Telescope is operated by the Joint
Astronomy Centre on behalf of Particle Physics and Astronomy
Research Council (PPARC). TRG is supported by the Gemini
Observatory, which is operated by the Association of Universities
for Research in Astronomy, Inc., on behalf of the international
Gemini partnership of Argentina, Australia, Brazil, Canada, Chile,
the United Kingdom and the United States of America.  This work was
partially supported by a PPARC visitors grants from PPARC and the
Royal Society. YP's studies are partially supported by a Small
Research Grant from American Astronomical Society. This research
has made use of the SIMBAD database, operated at CDS, Strasbourg,
France.

\end{document}